\documentclass[a4paper,12pt]{article}

\usepackage{graphicx}
\usepackage{array}
\usepackage{amsmath}
\usepackage{amssymb}
\usepackage{latexsym}
\usepackage{subfigure}
\usepackage{hyperref}
\usepackage{ulem}
\usepackage{color}
\renewcommand\sout{\bgroup \color{red} \ULdepth=-.5ex \ULset}

\newcommand {\beq}{\begin{eqnarray}}
\newcommand {\eeq}{\end{eqnarray}}
\newcommand {\non}{\nonumber\\}

\newcommand {\gto}{\stackrel{g}{\to}}

\newcommand {\1}[1]{\frac{1}{#1}}

\newcommand {\ph}{\varphi}

\newcommand {\Ph}{\Phi}
\newcommand {\Phd}{\Phi^{\dagger}}

\newcommand {\del}{\partial}
\newcommand {\dagg}{^{\dagger}}
\newcommand {\pri}{^{\prime}}

\newcommand {\tr}{{\rm tr}\,}
\newcommand {\GC}{G^{\mathbb C}}
\newcommand {\HC}{H^{\mathbb C}}

\makeatletter
\@addtoreset{equation}{section}
\makeatother

\date{empty}
\begin{document}
\begin{titlepage}
\null
\begin{flushright}
\end{flushright}
\vskip 0.5cm
\begin{center}
{\Large \bf BPS pion domain walls \\
in the supersymmetric chiral Lagrangian
}
\vskip 1.1cm
\normalsize
\renewcommand\thefootnote{\alph{footnote}}

{\large
Sven Bjarke Gudnason$^{\dagger}$\footnote{bjarke(at)impcas.ac.cn},
Muneto Nitta$^{\ddagger}$\footnote{nitta(at)phys-h.keio.ac.jp}
and Shin Sasaki$^{\P}$\footnote{shin-s(at)kitasato-u.ac.jp}
}
\vskip 0.7cm
  {\it
$^\dagger$ Institute of Modern Physics, Chinese Academy of Sciences,
  Lanzhou 730000, China\\
  $^\ddagger$ 
Department of Physics, and Research and Education Center for Natural Sciences, \\
\vskip -0.2cm
Keio University, Hiyoshi 4-1-1, Yokohama, Kanagawa 223-8521, Japan  
\vskip 0.1cm
$^{\P}$
  Department of Physics,  Kitasato University \\
  \vskip -0.2cm
  Sagamihara 252-0373, Japan
}
\vskip 0.5cm
\begin{abstract}
We construct exact solutions of 
BPS pion domain walls in 
the four-dimensional $\mathcal{N}=1$ supersymmetric $SU(N)$ chiral
Lagrangian with pion masses introduced via linear and quadratic
superpotentials. 
The model admits $N$ discrete vacua in the center of $SU(N)$ for the
linear superpotential. 
In addition to the latter, new vacua appear for the quadratic
superpotential. 
We find that the domain wall solutions of pions (Nambu-Goldstone
bosons) that interpolate
between a pair of (pion) vacua preserve half of
supersymmetry.
Contrary to our expectations, we have not been able to find domain
walls involving the quasi-Nambu-Goldstone bosons present in the theory, 
which in turn has the consequence that not all vacua of the theory are
connected by a BPS domain wall solution. 
\end{abstract}
\end{center}

\end{titlepage}

\newpage
\setcounter{footnote}{0}
\renewcommand\thefootnote{\arabic{footnote}}
\pagenumbering{arabic}

\tableofcontents

\section{Introduction}

Domain walls that separate two vacua are topological defects appearing 
in various subjects of physics from condensed matter physics to field
theory, high energy physics \cite{Vachaspati:2006zz}, QCD
\cite{Eto:2013hoa}, and cosmology \cite{Vilenkin:2000}. 
In supersymmetric theories, 
Bogomol'nyi-Prasad-Sommerfield (BPS) domain walls 
are the most stable configurations,
studied extensively in the literature,
such as supergravity \cite{Cvetic:1991vp} 
and ${\cal N}=1$ supersymmetric QCD \cite{Dvali:1996xe}.
They preserve half of supersymmetry
(therefore called 1/2 BPS states) and their tension is given by the
central charge in 1+1 dimensions \cite{Witten:1978mh}.
In 3+1 dimensions the tension of the 1/2 BPS domain walls coincides
instead with a tensorial charge present only in theories with broken
translational invariance \cite{Dvali:1996xe}.
Domain walls were also studied in 
theories with extended supersymmetry such as 
${\cal N}=2$ supersymmetric hyper-K\"ahler sigma models
\cite{Abraham:1992vb}  
and ${\cal N}=2$ supersymmetric 
Abelian \cite{Tong:2002hi} 
and non-Abelian \cite{Isozumi:2004jc} gauge theories.
If multiple domain walls with different angles join
at a junction, the total configuration is a 1/4 BPS state 
preserving a quarter of supersymmetry  
both in ${\cal N}=1$ \cite{Gibbons:1999np,Oda:1999az} 
and ${\cal N}=2$ \cite{Eto:2005cp} 
supersymmetric gauge theories.
See Refs.~\cite{Tong:2005un,Eto:2006pg,Shifman:2007ce}
for reviews.

In this paper, we study BPS pion domain walls 
in the ${\cal N}=1$ supersymmetric chiral Lagrangian 
with pion mass terms.
The model appears 
as the low-energy effective theory 
of supersymmetric QCD in 
supersymmetric vacua with broken chiral symmetry. 
The $SU(N)$ chiral Lagrangian with the simplest 
pion mass term admits $N$ symmetric discrete vacua 
in the center elements of $SU(N)$. 
We construct exact solutions of 
BPS $SU(2K)$
pion domain walls interpolating between 
the pion vacua present in the theory
and find that these domain walls carry 
$SU(2K)/[SU(K) \times SU(K) \times U(1)]$
orientational moduli 
as well as translational moduli.
These domain walls are special solutions interpolating only 2 of $2K$
vacua. We have not been able to find any domain wall solutions
connecting any of the other $2K-2$ vacua.
We construct the low-energy effective field theory 
on the domain wall for the $N=2$ case 
and obtain the $\mathbb{C}P^1$ model.
This case is similar to the moduli space found for vortices in
$U(2)$ gauge theories \cite{Hanany:2003hp}.
The $SU(2)$ case ($N=2$) reduces to 
a domain wall in the $O(4)$ model admitting 
two discrete vacua \cite{Kudryavtsev:1999zm,Nitta:2012wi}, in which 
the ${\mathbb C}P^1$ moduli of the domain wall were already found. 
In contrast to pion domain walls in non-supersymmetric theories  
\cite{Hatsuda:1986nq,Eto:2013hoa} that are topologically and
dynamically unstable,  
pion domain walls found in this paper saturate the BPS bound 
and are therefore stable classically and quantum mechanically  
(even non-perturbatively).

In supersymmetric theories,
a global symmetry $G$ is extended to its 
complex extension $\GC$ since 
the potential is constructed from 
a superpotential which is holomorphic 
in the chiral superfields.
Consequently, 
spontaneously broken global symmetry 
in supersymmetric vacua 
results in additional massless bosons, 
called quasi-Nambu-Goldstone (NG) bosons 
\cite{Kugo:1983ma,Lerche:1983qa}  
in addition to the usual NG bosons.
These massless bosons 
together with their fermionic superpartners, called 
quasi-NG fermions \cite{Buchmuller:1982xn},
constitute chiral multiplets.
The NG and quasi-NG bosons  
must parametrize a K\"ahler manifold
as required from supersymmetric nonlinear sigma models
\cite{Zumino:1979et}.  
The general framework to construct
low-energy effective theories
was provided in Refs.~\cite{Bando:1983ab}.  
In the case of chiral symmetry breaking 
$SU(N)_{\rm L}\times SU(N)_{\rm R} \to SU(N)_{\rm L+R}$,  
there must appear the same number of 
quasi-NG bosons as the number of NG bosons 
(pions) and the target space is
$SU(N)^{\mathbb{C}}\simeq SL(N,{\mathbb{C}})$ \cite{Kotcheff:1988ji}. 
The most general K\"ahler potential 
is an arbitrary function of 
$G$-invariants,
corresponding to the deformation of  
directions of quasi-NG bosons, 
which cannot be fixed by $G$ 
\cite{Kotcheff:1988ji,Shore:1988mn,Higashijima:1997ph,Nitta:1998qp}.
Manifestly supersymmetric 
higher-derivative corrections have recently been 
constructed, including the example of chiral symmetry breaking
\cite{Nitta:2014fca}.
A supersymmetric Skyrme term has been constructed recently 
\cite{Gudnason:2015ryh} 
but the usual kinetic term canceled out 
as in the case of baby (lower dimensional) Skyrmions 
\cite{Adam:2013awa}. 
In this paper, we study -- for chiral symmetry breaking --
supersymmetric pion mass terms  
preserving the vector symmetry $H = SU(N)_{\rm L+R}$.
In the case of the simplest superpotential, 
the potential admits $N$ symmetric discrete vacua 
for the $SU(N)$ case.
We construct BPS pion domain walls interpolating between 
the pion vacua of the theory.
These vacua for which we are able to find domain wall
solutions are antipodal points on the target space. 
However, as we mentioned, not all the supersymmetric vacua are
connected by domain walls; vacua with an imaginary part require
quasi-NG bosons to be turned on.
For this type of domain wall -- although we have found the BPS
equations -- we have not been able to find a domain wall solution,
neither analytically nor numerically.
Using an appropriate Ansatz, we have reduced the BPS matrix equation
to a complex scalar equation which describes one NG mode and one
quasi-NG mode, for which we can show that no solutions exist. Although
we do not yet have a solid proof of absence of the remaining domain
wall, our results provide some evidence.

As a similar model,
the (non-supersymmetric) $U(N)$ chiral Lagrangian 
with the pion mass term 
admits a non-Abelian sine-Gordon soliton 
that carries ${\mathbb C}P^{N-1}$ moduli \cite{Nitta:2014rxa}.
The low-energy effective theory on said domain wall is given by  
the ${\mathbb C}P^{N-1}$ model \cite{Eto:2015uqa}.
Such a $U(N)$ chiral Lagrangian appears e.g.~in the Josephson junction 
of two non-Abelian superconductors, 
in which a non-Abelian sine-Gordon soliton 
describes a non-Abelian vortex (color-magnetic flux tube) 
from the bulk point of view 
\cite{Nitta:2015mma}, 
that is a non-Abelian extension of Josephson 
vortices in field theory \cite{Nitta:2012xq}.
For the non-Abelian sine-Gordon soliton 
in the $U(N)$ chiral Lagrangian,  
one has to consider 
the group $U(N)$ instead of $SU(N)$.
We do not need a $U(1)$ part and
consider instead the simple group $SU(N)$. 
Consequently, our configurations
separate into two {\it different} vacua 
so they are domain walls, 
but two spatial infinities of a sine-Gordon soliton 
are in the {\it same} vacuum.
We also show that there is no BPS domain wall interpolating the
same vacuum in our model. Only two physically distinct vacua can be
connected by a BPS pion domain wall. As a consequence, we find no
domain wall solutions for the $SU(2K+1)$ case.

This paper is organized as follows.
In Sec.~\ref{sec:susy-nlr}, we give a brief review 
of the 
supersymmetric nonlinear sigma model 
and chiral symmetry breaking 
in supersymmetric theories, 
and discuss supersymmetric pion mass terms.
In Sec.~\ref{sec:BPS-wall}, 
we construct non-Abelian BPS domain walls.
In Sec.~\ref{sec:eff_th}, we construct 
the effective theory on the domain wall 
which is the ${\mathbb C}P^{1}$ model.
Section \ref{sec:conc} is devoted to a summary as well as
discussions. 
We use the notation of the textbook of 
Wess and Bagger \cite{Wess:1992cp}.

\section{Supersymmetric chiral Lagrangian}
\label{sec:susy-nlr}

Subsections \ref{sec:snlsm} and \ref{sec:chisb} 
are devoted to a review of supersymmetric 
nonlinear sigma models and chiral symmetry breaking 
in supersymmetric theories, respectively, while 
the supersymmetric mass term 
in Subsection \ref{sec:susy-mass} 
has not been discussed in the literature.

\subsection{Supersymmetric nonlinear sigma models}
\label{sec:snlsm}

In four-dimensional $\mathcal{N} = 1$ supersymmetric theories, we have
$N$ chiral superfields $\Phi^i$, $(i=1, \ldots, N)$ whose component
expansion in the chiral basis,
$y^m = x^m + i \theta \sigma^m \bar{\theta}$, is given by 
\begin{align}
\Phi^i (y,\theta) = \varphi^i (y) 
+ \theta \psi^i (y) + \theta^2 F^i(y),
\end{align}
where $\varphi^i$ is a complex scalar field, $\psi^i$ is a Weyl
fermion and $F^i$ is a complex auxiliary field. 
The supersymmetric Lagrangian is described by a K\"ahler potential
$K(\Phi,\Phi\dagg)$ as well as a superpotential $W(\Phi)$, where the
first is a function of the superfields, $\Phi^i$, and the latter is a
holomorphic function
\begin{align}    
\mathcal{L}
  &= \int d^4\theta\; K(\Ph,\Phd) 
+ \left(\int d^2 \theta \; W(\Ph) + {\rm c.c.}\right) \non
&= - g_{i\bar{\jmath}}(\ph,\bar{\ph})
\del_m \ph^i \del^m \bar{\ph}^{\bar{\jmath}}
+ g_{i\bar{\jmath}}(\ph,\bar{\ph})
F^i F^{*\bar{\jmath}} + F^i \frac{\del W}{\del \ph^i} + F^{*\bar{\jmath}}
\frac{\del W^*}{\del \bar{\ph}^{\bar{\jmath}}}
     + \mbox{(fermion terms)},
\label{eq:kinetic_term}
\end{align}
where 
$g_{i\bar{\jmath}} \equiv \frac{\partial}{\partial \varphi^i}
\frac{\partial}{\partial \bar{\varphi}^{\bar{\jmath}}} K(\varphi, \bar{\varphi})$
is the K\"ahler metric.
The potential $V$ can be written in terms of the superpotential as
\beq
V = g_{i\bar{\jmath}} F^i F^{*\bar{\jmath}} 
 = g^{i\bar{\jmath}} \frac{\del W}{\del\ph^i}\frac{\del W^*}{\del\bar{\ph}^{\bar{\jmath}}},
\eeq
while the auxiliary field is solved by
\beq
F^i = -g^{i\bar{\jmath}} \frac{\del W^*}{\del\bar{\ph}^{\bar{\jmath}}}.
\eeq
Here $g^{i\bar{\jmath}}$ is the inverse of the K\"ahler metric
$g_{i\bar{\jmath}}$. 
The $G$-invariance of the K\"ahler potential implies that the
following transformation
\beq
 K(\Phi,\Phi\dagg) \gto K(\Phi\pri,\Phi^{\prime\dagger}) =
  K(\Phi,\Phi\dagg) + F(\Phi,g) + F^*(\Phi\dagg,g), 
  \label{eq:G-transf.}
\eeq
is preserved; i.e.~the transformation with $F$ ($F^*$) being a
(n anti-)holomorphic function of $\Phi$ ($\Phi^\dag$) which are
determined by a group element $g \in G$. This transformation is called
a K\"ahler transformation and the latter two terms in the above
equation vanish under the superspace integral $\int d^4 \theta$.

\subsection{Supersymmetric chiral Lagrangian}
\label{sec:chisb}

Let us now consider chiral symmetry breaking of the form
\begin{align}
 G = SU(N)_{\rm L} \times SU(N)_{\rm R} 
\to H = SU(N)_{\rm L+R}.
\end{align}
The NG modes corresponding to the above symmetry breaking span the
following coset space 
\beq
G/H = \frac{SU(N)_{\rm L} \times SU(N)_{\rm R}}{SU(N)_{\rm L+R}}
\simeq 
SU(N).
\eeq
We denote the generators of the coset by
$T_A \in\mathfrak{su}(N)$, which take value in the $SU(N)$ algebra. 
It was shown in Ref.~\cite{Lerche:1983qa} that when the vacuum
expectation value (VEV) giving rise to the symmetry breaking
belongs to a real representation of $SU(N)$, then the number of
quasi-NG boson is exactly the same as the number of NG bosons; this is
also called a maximal realization.

Chiral symmetry breaking belongs to said class and the total target
space is given by
\beq 
\GC/\HC \simeq 
SU(N)^{\mathbb C} 
\simeq
 SL(N,{\mathbb C}) \simeq T^* SU(N).
 \eeq
The NG supermultiplet is expressed as the following coset
representative 
\begin{align}
 M = \exp (i \Phi^i T_A \delta_i^A) 
\in \GC/\HC, 
\label{eq:pion_superfield}
\end{align}
where the NG superfields take the form
\begin{align}
\Phi^i (y,\theta) 
= \pi^i (y) + i \sigma^i (y) 
+ \theta \psi^i (y) + \theta\theta F^i(y), 
\end{align}
with $\pi^i$ being NG bosons, $\sigma^i$ quasi-NG bosons -- both of
which are real fields -- and finally $\psi^i$ quasi-NG fermions. 
The NG supermultiplets obey the following nonlinear transformation law
\beq
M \to M' = g_{\rm L} M g_{\rm R}^\dag, \qquad
(g_{\rm L}, g_{\rm R}) 
\in SU(N)_{\rm L} \times SU(N)_{\rm R}.
\eeq
In the vacuum $M = {\bf 1}_N$, 
the unbroken symmetry $H= SU(N)_{\rm L+R}$
defined by $g_L=g_R$ remains.\footnote{ 
For chiral symmetry breaking 
in supersymmetric vacua, 
the unbroken group $H = SU(N)_{\rm L+R}$ 
is not unique, and is 
further broken to a subgroup
when some quasi-NG bosons get VEVs
\cite{Kotcheff:1988ji},
where some of the quasi-NG bosons change to NG bosons 
\cite{Kotcheff:1988ji,Nitta:1998qp}.
}
From the following transformation
\beq
  M M^\dagger \to g_{\rm L}  M M^\dagger g_{\rm L}^\dagger,
\eeq
the simplest K\"ahler potential, that is invariant under the
$SU(N)_{\rm L}\times SU(N)_{\rm R}$ symmetry, is just
\beq
K_0 =  f_{\pi}^2 \tr (M M^\dagger),\label{eq:chiral-Kahler0}
\eeq
where $f_{\pi}$ is a constant.
The bosonic part of the Lagrangian -- corresponding to the above
K\"ahler potential -- to leading order in the derivative expansion is 
\beq
\mathcal{L}_0 = - f_{\pi}^2 \tr (\del_m M \del^m M^\dagger),
\eeq
where $M$ is the lowest component of the NG superfield given in
Eq.~\eqref{eq:pion_superfield}.

From the left-invariant Maurer-Cartan one-form 
$i M^{-1} \frac{\del M}{\del \ph^i }$ we define 
the holomorphic vielbein $E^A_i(\ph)$ and their conjugates 
as
\beq
 i M^{-1} \frac{\del M}{\del \ph^i}= E^A_i(\ph) T_A,
\quad 
- i \frac{\del M^\dag}{\del\bar{\ph}^{\bar{\imath}}} M^{-1\dag}
= E^{*\bar{A}}_{\bar{\imath}}(\bar{\ph}) T_{\bar{A}}.
\eeq
Their pull-backs to space-time give
\beq
 i M^{-1} \del_m M = E^A_i(\ph) T_A \del_m \ph^i, \qquad
 -i (\del_m M^\dagger) M^{-1\dagger}  
= E^{*\bar{A}}_{\bar{\imath}}(\bar{\ph}) T_{\bar{A}} \del_m \bar{\ph}^{\bar{\imath}}.
\eeq
By using the vielbein, the Lagrangian for the bosonic fields can be
rewritten as 
\beq
 {\cal L}_0 &=& - f_\pi^2 \tr (  M T_A T_{\bar{B}} M^\dag  )  
 E^A_i(\ph)  E^{*\bar{B}}_{\bar{\jmath}}(\bar{\ph}) \del_m \ph^i
 \del^m \bar{\ph}^{\bar{\jmath}}
 = - G_{A\bar{B}} E^A_i(\ph)  E^{*\bar{B}}_j(\bar{\ph})
 \del_m \ph^i \del^m \bar{\ph}^{\bar{\jmath}} \non
&=& - g_{i\bar{\jmath}}(\ph,\bar{\ph}) \del_m \ph^i \del^m \bar{\ph}^{\bar{\jmath}},
\eeq
with the K\"ahler metric $g_{i\bar{\jmath}}$ and 
the metric $G_{A\bar{B}}$ on the tangent space, defined by 
\beq
 g_{i\bar{\jmath}}(\ph,\bar{\ph}) 
 &=&  f_\pi^2 \tr  (  M T_A T_{\bar{B}} M^\dag  )  E^A_i(\ph)
 E^{*\bar{B}}_{\bar{\jmath}}(\bar{\ph}) 
 = G_{A\bar{B}} E^A_i(\ph)  E^{*\bar{B}}_{\bar{\jmath}}(\bar{\ph}) ,\\
 G_{A\bar{B}} &=& f_\pi^2 \tr  (  M T_A T_{\bar{B}} M^\dag  ),
\eeq
respectively.

The K\"ahler potential in Eq.~(\ref{eq:chiral-Kahler0}) is the
simplest one, while the most general K\"ahler potential can be written 
as \cite{Kotcheff:1988ji,Nitta:1998qp}
\beq
 K = f(\tr (M M^\dagger),  \tr [(M M^\dagger)^2], \cdots, 
\tr [(M M^\dagger)^{N-1}]),  \label{eq:chiral-Kahler-general}
\eeq
with an {\it arbitrary} function of $N-1$ arguments.

If we set all quasi-NG bosons to zero
\cite{Shore:1988mn,Kotcheff:1988ji} 
\beq
 U = M|_{\sigma^i=0} \in SU(N), \label{eq:QNGzero}
\eeq
we get the $SU(N)$ chiral Lagrangian
\beq
\mathcal{L} = - f_\pi^2 \tr (\del_m U \del^m  U^\dagger)
= f_\pi^2 \tr (U^\dagger \del_m U)^2,
\eeq
where the decay constant $f_\pi$ is determined from the function $f$. 
Here, we have used that
\beq
 G_{A\bar{B}}|_{\sigma=0} = f_\pi^2 \delta_{A\bar{B}},  \qquad
 E^A_i|_{\sigma^j=0} = e^A_i(\pi).
\eeq
with the normalization of generators
$\tr[T_A T_{\bar{B}}]=\delta_{A\bar{B}}$ 
and the vielbein $e^A_i(\pi)$ for $SU(N)$.

\subsection{Supersymmetric mass term}\label{sec:susy-mass}

The pion mass term in the chiral Lagrangian 
breaks the $G=SU(N)_{\rm L} \times SU(N)_{\rm R}$ symmetry 
explicitly. 
It is often considered that explicit breaking terms
do not break the vector symmetry $SU(N)_{\rm L+R}$. 
Here we consider such mass terms preserving 
the vector symmetry $SU(N)_{\rm L+R}$. 
The superpotential invariant under $SU(N)_{\rm L+R}$ 
is 
\beq
  W = g (\tr (M), \tr (M^2), \cdots ,  \tr (M^{N-1})),
\eeq
with an arbitrary function $g$ of $N-1$ arguments.

In this paper, we consider only functions of the trace $M$, for
simplicity: 
\beq
  W = w (\tr M),
\eeq
with an arbitrary function $w$.
The auxiliary fields are solved as
\beq
 F^i = -g^{i\bar{\jmath}} \frac{\del W^*}{\del\bar{\ph}^{\bar{\jmath}}}
 = -i \bar{w}'(\tr M^\dagger) \tr(M^\dagger T_{\bar{A}}) g^{i\bar{\jmath}}
 E^{*\bar{A}}_{\bar{\jmath}}(\bar{\ph}),  \label{eq:F}
\eeq
where the prime denotes differentiation with respect to the argument,
and so the potential term can be written as
\beq
 V = g_{i\bar{\jmath}} F^i F^{*\bar{\jmath}} 
 =  \left|w'(\tr M)\right|^2
    \tr(M^\dag T_{\bar{B}}) \tr(M T_A) G^{A\bar{B}}.
\eeq
Here $G^{A\bar{B}}$ is the inverse of the metric $G_{A\bar{B}}$ on the
tangent space. 
The supersymmetric vacua are given by
\beq
 F=0 \Leftrightarrow w'(\tr M)\, \tr(M T_A) = 0.
\eeq
In the next two subsections we will consider the two simplest
possibilities for a chiral symmetry breaking mass term, conserving
the vector symmetry $SU(N)_{\rm L+R}$.

\subsubsection{Linear superpotential}

The simplest superpotential 
\beq
 W = w(\tr M) = \frac{m}{N} \tr M,   \label{eq:linearpot}
\eeq
with mass $m\in\mathbb{R}$, admits $N$ symmetric supersymmetric vacua,
given by\footnote{A phase for the mass will just rotate all the
  supersymmetric vacua, so we can set the phase to zero without loss
  of generality. } 
\beq
 M = \omega^k {\bf 1}_N, \qquad
 k = 0,1,2,\cdots,N-1, \qquad 
 \omega \equiv \exp (2\pi i /N),
 \label{eq:linearpot_generalvacua}
\eeq
namely the center elements of $SL(N,{\mathbb C})$.

Let us point out a crucial fact about the restriction to the NG
subspace: $M=M|_{\sigma^i=0}=U\in SU(N)$.
In this subspace, we can write
\beq
\tr[U] = \tr\left[\exp\left\{i\theta^A T^A\right\}\right] \in \mathbb{R},
\eeq
if and only if $\theta^A\in\mathbb{R}$ are real parameters.
Since for the NG restriction $\theta^A$ are indeed real parameters,
the above expression holds.\footnote{To realize that the expression
  holds, it is enough to realize that an $i$ can only come from the
  product of an odd number of generators which is traceless and
  therefore does not contribute to the trace. All even powers of the
  generators have no $i$ and thus the trace is a real quantity. }
Therefore, in this subspace only the vacua
\beq
U = \mathbf{1}_N, \qquad
U = -\mathbf{1}_N,
\eeq
can be reached for even $N$ and only the vacuum 
\beq
U = \mathbf{1}_N,
\eeq
is possible for odd $N$. 
In order to reach the general $\omega^k\neq \pm\mathbf{1}_N$
vacua, we need to turn on the quasi-NG directions.

We note that for the $N=2$ case, the NG boson part of the Lagrangian
with the superpotential \eqref{eq:linearpot} reduces to the well-known $O(4)$ model:
\beq
\mathcal{L} = - f_\pi^2\del_m {\bf m} \cdot \del^m {\bf m},
\eeq
with ${\bf m}=(m_1,\cdots,m_4)$ with the constraint ${\bf m}^2=1$ 
and the potential \cite{Kudryavtsev:1999zm,Nitta:2012wi}, 
\beq 
V = \frac{m^2}{2f_\pi^2} (m_1^2+m_2^2+m_3^2)
  = \frac{m^2}{2f_\pi^2} (1-m_4^2),
\eeq
admitting two vacua $m_4 = \pm 1$.

\subsubsection{Quadratic superpotential}

We will also consider the next-simplest potential, i.e.~a quadratic
potential of the form
\beq
 W = w(\tr M) = \frac{m}{2N^2}\left(\tr M\right)^2, \label{eq:quadpot}
\eeq
such that the vacuum equation now reads
\beq
\tr(M)\tr(M T_A) = 0,
\eeq
which has both the old type of vacua
\beq
M = \omega^k\mathbf{1}_N, \qquad
\omega = \exp\frac{2\pi i}{N},
\eeq
as well as new vacua
\beq
\tr M = 0.
\eeq
These new vacua are sections of $SL(N,\mathbb{C})$ and probably
connected spaces, but not connected to the old type of vacua.

The $SU(2)$ case of $N=2$, i.e.~the NG subspace of the model, now
reduces to the $O(4)$ model with the following potential 
\beq 
V = \frac{m^2}{2f_\pi^2} m_4^2 (1 - m_4^2),
\eeq
admitting three vacua: $m_4=\pm 1$ and $m_4=0$.
Notice that the vacua $m_4=\pm 1$ are point-like on the space of
vacua, whereas the vacuum $m_4=0$ is the manifold $S^2$:
$m_1^2 + m_2^2 + m_3^2 = 1$.
The latter can be interpreted as vacuum moduli. 

The $m_4=0$ vacuum breaks the global $SU(2)$ symmetry to $U(1)$.

\section{BPS pion domain walls}\label{sec:BPS-wall}

\subsection{BPS equation and Bogomol'nyi bound 
for domain walls}

BPS equations are obtained by the condition that the supersymmetry
transformation of fermions vanish. The transformation law of the
fermions in the chiral multiplet is given by
\begin{align}
\delta \psi^i = i \sqrt{2} \sigma^m \bar{\xi} \partial_m \varphi^i +
 \sqrt{2} \xi F^i,
\end{align}
where $\xi$ and $\bar{\xi}$ are transformation parameters.
Assuming that the fields $\varphi^i$ depend only on the
$x^1$-direction and imposing the half-BPS condition
$i \sigma^1 \bar{\xi} = \xi$, we obtain the following BPS equation for
domain walls:
\beq
\del_1 \ph^i + F^i = 0.
\eeq
From Eq.~(\ref{eq:F}), the above equation reads
\beq
  \del_1 \ph^i  
  = i \bar{w}'(\tr M^\dag)\, \tr(M^\dag T_{\bar{A}})
  g^{i\bar{\jmath}} E^{*\bar{A}}_{\bar{\jmath}}.
\eeq
By multiplying by $E^B_i T_B$ on the both sides, we obtain 
the invariant form of the BPS equation 
\beq
  i M^{-1} \del_1 M = i \bar{w}'(\tr M^\dag) \tr(M^\dag T_{\bar{B}})
  T_A G^{A\bar{B}}.
  \label{eq:BPS-M}
\eeq
If we restrict to the NG-boson subspace, $M=M|_{\sigma^i=0}=U$, we get 
\beq
  i U^\dagger \del_1 U 
  = \frac{i}{f_\pi^2} \bar{w}'(\tr U^\dag) \tr(U^\dag T_A) T_A.
  \label{eq:BPS-U}
\eeq

The BPS equation \eqref{eq:BPS-M} can also be obtained from the
Bogomol'nyi bound.
The Lagrangian can be written as
\beq
\mathcal{L} =
- G_{A\bar{B}} E_i^A(\varphi) E_{\bar{\jmath}}^{*\bar{B}}(\bar{\varphi})
  \del_1\varphi^i \del^1\bar{\varphi}^{\bar{\jmath}}
- G^{A\bar{B}} |w'(\tr M)|^2 \tr(M T_A) \tr(M^\dag T_{\bar{B}}),
\eeq
yielding the energy for domain walls
\begin{align}
  E &= \int dx^1 \left(G_{A\bar{B}} E_i^A(\varphi)
  E_{\bar{\jmath}}^{*\bar{B}}(\bar{\varphi})
  \del_1\varphi^i \del^1\bar{\varphi}^{\bar{\jmath}}
  + G^{A\bar{B}} |w'(\tr M)|^2 \tr(M T_A) \tr(M^\dag T_{\bar{B}})
  \right) \non
    &=\int dx^1\; G_{A\bar{B}} \left[
    E_i^A(\varphi)\del_1\varphi^i
    - i G^{A\bar{C}} \bar{w}'(\tr M^\dag) \tr(M^\dag T_{\bar{C}})\right]\non
  &\phantom{=\int dx^1\; \ } \times
  \left[
    E_{\bar{\jmath}}^{*\bar{B}}(\bar{\varphi})\del_1\bar{\varphi}^{\bar{\jmath}}
    + i G^{D\bar{B}} w'(\tr M) \tr(M T_D)\right] + T,
\end{align}
where the domain wall topological charge is defined by
\begin{align}
  T &\equiv \int dx^1 \left(-i E_i^A(\varphi)\del_1\varphi^i
  w'(\tr M) \tr(M T_A)
  + i E_{\bar{\jmath}}^{*\bar{B}}(\bar{\varphi})
  \del_1\bar{\varphi}^{\bar{\jmath}}
  \bar{w}'(\tr M^\dag) \tr(M^\dag T_{\bar{B}})\right) \non
  &= \int dx^1 \left(
  w'(\tr M) \tr(\del_1 M) + \bar{w}'(\tr M^\dag) \tr(\del_1 M^\dag)
  \right)\non
  &= |[2\Re(W)]_{x=-\infty}^{x=+\infty}|. \label{eq:DWtension}
\end{align}
If we now consider the restriction to the NG subspace (i.e.~setting
$\sigma^i=0$), then we get the energy for the NG domain walls
\begin{align}
E &= \int dx^1\; \left[f_\pi^2 \tr (i U^\dagger \del_1 U)^2
 + f_\pi^{-2} |w'(\tr U)|^2 \tr(U T_A)\tr (U^\dagger T_A)\right] \non
 &= \int dx^1\; \tr\left[\left(f_\pi U^\dagger \del_1 U
   - f_\pi^{-1} w'(\tr U)\tr(U T_A)T_A\right) \label{eq:Ewall_line2}
   \left(f_\pi \del_1 U^\dagger U
   - f_\pi^{-1} \bar{w}'(\tr U^\dag)\tr (U^\dag T_B)T_B\right)\right] \non
   &\phantom{=\ }
   + T, 
\end{align}
in turn reproducing the BPS equation for the NG subspace
\eqref{eq:BPS-U} and the domain wall topological charge $T$ is now
given by 
\beq
T = \int dx^1 \left(w'(\tr U) \tr (\del_1 U)
    + \bar{w}'(\tr U^\dag) \tr (\del_1 U^\dag)\right)
= \big|[2\Re(W)]^{x=+\infty}_{x=-\infty} \big|.
\eeq
The energy $E$ is most severely bounded from below by $|T|$.
The bound is saturated when the quantity in the parentheses in
Eq.~\eqref{eq:Ewall_line2} vanishes. This condition is nothing but the
BPS equation \eqref{eq:BPS-U}.

\subsection{Linear superpotential}

In this subsection we consider the simplest superpotential, namely the
linear one of Eq.~\eqref{eq:linearpot}. 
In this case, the BPS equation reads
\beq
 i M^{-1}\del_1 M =  \frac{i m}{N}\,\tr(M^\dag T_{\bar{B}}) T_A G^{A\bar{B}}.
\eeq
Restricting to the NG-boson subspace, $M=M|_{\sigma^i=0}=U$, we get
\beq
  i U^\dag \del_1 U 
  = \frac{i m}{N f_\pi^2} \tr(U^\dag T_A) T_A, \label{eq:BPS-U-linear}
\eeq
where we have used the expression of the inverse metric on the tangent 
space: $G^{A\bar{B}}=f_\pi^{-2}\delta^{A\bar{B}}$. 

With this superpotential, we can calculate the tension of the domain
wall using Eq.~\eqref{eq:DWtension}, which for the vacua
\eqref{eq:linearpot_generalvacua} gives 
\beq
T_k = 2m |\Re(\omega^k) - 1| = 4m\sin^2\frac{\pi k}{N}, \qquad
k \in \mathbb{Z}, \label{eq:tensionk}
\eeq
where we have assumed that the domain wall starts from the vacuum
$M=\mathbf{1}_N$ and goes to the vacuum $M=\omega^k\mathbf{1}_N$.
The fundamental domain wall, i.e.~interpolating between two nearest
vacua, thus has the tension
\beq
T_1 = 2m |\Re(\omega) - 1| = 4m\sin^2\frac{\pi}{N}.
\label{eq:tension1}
\eeq
A domain wall with the maximum tension is given by 
\beq
  \frac{k}{N} =\1{2}  \;\; && \mbox{for even }\; N , \non
 \frac{k}{N \pm 1} =\1{2}  \;\; &&  \mbox{for odd }\; N.
\eeq

If we now restrict to the NG subspace, $M=M|_{\sigma^i=0}=U$, then
only real vacua exists and thus the single domain wall exists only for
even $N$ and interpolates between $U=\mathbf{1}_N$ and
$U=-\mathbf{1}_N$, giving the domain wall tension
\beq
T = 4m\sin^2\frac{\pi}{2} = 4m.
\eeq
A double domain wall for even $N$ or a single domain wall for odd $N$
would wind $2\pi$ and thus have a vanishing tension.
Since the superpotential is not double valued, these solutions do not
exist. 
Alternatively, we can think of two domain walls in the NG subspace for
even $N$ as a domain wall and an anti-domain wall, which thus have
zero overall topological charge.
They may exist locally if well separated, but they are likely to
decay to the vacuum, i.e.~to the trivial topological sector.

\subsubsection{$SU(2)$ solution}\label{sec:SU2sol}

We will begin with the simplest possible solution, which is in the NG
subspace and for $N=2$; namely the $SU(2)$ case.
The linear superpotential \eqref{eq:linearpot} gives rise to two
discrete vacua $U = \pm {\bf 1}_2$.
The general element of $SU(2)$ can be written as 
\beq
 U = \exp \left(i \frac{\theta}{2} {\bf n}\cdot \sigma\right)
 = \cos\frac{\theta}{2}   {\bf 1}_2
 + i {\bf n}\cdot \sigma \sin\frac{\theta}{2},
  \label{eq:SU(2)elem}
\eeq
with a unit vector 
${\bf n}=(n_1,n_2,n_3)$, (${\bf n}^2=1$) and 
the Pauli matrices $\sigma_A$. 
We construct a domain wall interpolating between 
$U={\bf 1}_2$, ($\theta =0$) at $x\to+\infty$
and $U=-{\bf 1}_2$, ($\theta = 2\pi$) at $x\to -\infty$.
By using an $SU(2)$ transformation, 
Eq.~\eqref{eq:SU(2)elem} can be diagonalized 
without loss of generality 
to ${\bf n}=(0,0,1)$, yielding:
\beq
 U_0 = {\rm diag}\, (e^{i\theta/2}, e^{-i\theta/2}).
\label{eq:SU2_solution}
\eeq
Then, the BPS equation \eqref{eq:BPS-U-linear} reduces to
\beq
\del_1\theta = -\frac{m}{f_\pi^2} \sin\frac{\theta}{2},
\label{eq:SG}
\eeq
which is the BPS equation for the sine-Gordon soliton.
A single soliton solution is
\beq
\theta(x^1) = 4 \arctan \exp\left[-\frac{m}{2f_\pi^2} (x^1-X)\right],
\label{eq:single}
\eeq
with the constant $X\in\mathbb{R}$ corresponding to the position of  
the soliton.
We thus find that the most general single soliton solution 
is Eq.~\eqref{eq:SU(2)elem} with Eq.~\eqref{eq:single}.
The general solution therefore has 
the moduli 
\beq 
  S^2 \simeq \frac{SO(3)}{SO(2)} 
 \simeq {\mathbb C}P^1 \simeq \frac{SU(2)}{U(1)},
\eeq
characterized by ${\bf n}$.
The tension for this domain wall is $T=4m$.

\subsubsection{$SU(2K)$ solutions}

In this section we consider the NG subspace for even $N=2K$, with 
$K\in\mathbb{Z}$. 
We now choose an Ansatz for the element $U$ for a single domain wall
as 
\begin{align}
& U_0 = {\rm diag}\, \left(
\exp \left(\frac{i\theta}{2} \right), \cdots,
\exp \left(\frac{i\theta}{2} \right), 
\exp \left(-\frac{i\theta}{2} \right), \cdots,
\exp \left(-\frac{i\theta}{2} \right) \right)
= \exp (i\theta T_0),    \label{eq:M0_2K} \\
& T_0 \equiv {\rm diag}\,
   \left(\frac{1}{2},\cdots,\frac{1}{2},-\frac{1}{2},\cdots,-\frac{1}{2}\right) .
\end{align}
The boundary conditions of $\theta$ for the domain wall are:
$\theta=0$, ($U={\bf 1}_{2K}$) at $x \to +\infty$
and $\theta=2\pi$, ($U=-{\bf 1}_{2K}$) at $x \to -\infty$.

The BPS equation \eqref{eq:BPS-U-linear} can now readily be calculated
as 
\begin{align}
  i\del_1\theta T_0 &=
  \frac{m}{2K f_\pi^2} \tr\left[U_0^\dag T_A\right]T_A\non
  &= \frac{m}{2K f_\pi^2}\sum_{k=K+1}^{2K}
  \frac{1}{k(k-1)}\left[K e^{-\frac{i\theta}{2}} + (k-1-K) e^{\frac{i\theta}{2}}
    - (k-1)e^{\frac{i\theta}{2}}\right] \non
  &\phantom{=\ }
       \times{\rm diag}
    \left(\underbrace{1, \cdots, 1}_{k-1}, 1-k, \underbrace{0,\cdots,0}_{2K-k}
    \right)\non
  &= -\frac{i m}{f_\pi^2} \sin\frac{\theta}{2} \sum_{k=K+1}^{2K}
  \frac{1}{k(k-1)}\; {\rm diag}
    \left(\underbrace{1, \cdots, 1}_{k-1}, 1-k, \underbrace{0,\cdots,0}_{2K-k}
    \right)\non
  &= -\frac{i m}{K f_\pi^2} T_0 \sin\frac{\theta}{2}.
\end{align}
The solutions are thus given by Eq.~\eqref{eq:single} with $m\to m/K$. 

Since there are only two real vacua, this is the general single domain
wall in the restricted NG subspace.
The tension is again $4m$.

The solution has the moduli
\beq
\frac{SU(2K)}{SU(K)\times SU(K)\times U(1)},
\eeq
in addition to the translational modulus.

\subsubsection{$SU(2)$ double domain wall case}

In this section we consider the NG subspace for the $N=2$ case, with a
double domain wall, interpolating from $\mathbf{1}_2$ back to
$\mathbf{1}_2$. 
We now choose an Ansatz for the element $U$ for a single domain wall
as 
\begin{align}
  & U_0 = {\rm diag}\, \left(\exp \left(i\theta\right),
  \exp \left(-i\theta \right) \right)
= \exp (i\theta T_0),  \\
& T_0 \equiv {\rm diag}\,
   \left(1,-1\right).
\end{align}
The boundary conditions of $\theta$ for the domain wall are:
$\theta=0$, ($U={\bf 1}_{2}$) at $x \to +\infty$ 
and $\theta=2\pi$, ($U={\bf 1}_{2}$) at $x \to -\infty$.

The BPS equation \eqref{eq:BPS-U-linear} now reads
\beq
\del_1\theta = -\frac{m}{2f_\pi^2} \sin\theta.
\label{eq:doubleDW}
\eeq
$\theta$ can interpolate from $\pi$ to $0$, which is the normal domain
wall solution of Sec.~\ref{sec:SU2sol} or from $\pi$ to $2\pi$, which
is simply the anti-domain wall solution (mod $2\pi$).
Due to the fact that the right-hand side of Eq.~\eqref{eq:doubleDW} is
negative (positive) semi-definite for $\theta$ in the range $[0,\pi]$
($[\pi,2\pi]$), no BPS pion domain wall solution (i.e.~NG boson
domain wall) can interpolate between $2\pi$ and $0$.

This result extends trivially to $SU(2K)$ and since $SU(2K+1)$ only
has the single vacuum of the double domain wall, also no BPS solutions
exist for odd $N=2K+1$.

\subsubsection{$SL(3,\mathbb{C})$ case}

We now attempt to relax the restriction to the NG subspace, which is a
necessity if we are to consider the domain wall between the general
vacua $M=\mathbf{1}_N$ and $M=\omega\mathbf{1}_N$.
We will start by considering $SL(3,\mathbb{C})$.
We first diagonalize an $SL(3,\mathbb{C})$ element $M$ as
\beq
&& M_0 = {\rm diag}\, \left(
\exp \left(\frac{i\theta}{3} \right),
\exp \left(\frac{i\theta}{3} \right), 
\exp \left(-\frac{i2\theta}{3} \right) \right)
= \exp (i\theta T_0),    \label{eq:M0_3} \\
&& T_0 \equiv {\rm diag}\,
   \left(\frac{1}{3},\frac{1}{3},-\frac{2}{3}\right) .
\eeq
We consider the boundary conditions of $\theta$ 
for a domain wall:
$\theta=0$, ($M={\bf 1}_3$) at $x \to +\infty$
and $\theta=2\pi$, ($M=\omega {\bf 1}_3$) at $x \to -\infty$.

Now the situation is a little more complicated because when we are not
restricting to the NG subspace, we need also to take into account the
metric on the tangent space $G_{A\bar{B}}$.
The BPS equation \eqref{eq:BPS-M} now reads
\beq
i\del_1\theta = \frac{m e^{\frac{i2\theta}{3}}
  \left(1 - e^{i\bar{\theta}}\right)}
{f_\pi^2\left(e^{i\theta} + 2e^{i\bar{\theta}}\right)},
\label{eq:BPS_SL3}
\eeq
where we have used the inverse metric on the tangent space
\begin{align}
  &G^{1\bar{1}} = G^{2\bar{2}} = G^{3\bar{3}}
  = \frac{1}{f_\pi^2} e^{2\Im(\theta)/3},\\
  &G^{4\bar{4}} = G^{5\bar{5}} = G^{6\bar{6}} = G^{7\bar{7}}
  = \frac{1}{2f_\pi^2}e^{-4\Im(\theta)/3} +
  \frac{1}{2f_\pi^2}e^{2\Im(\theta)/3},\\
  &G^{4\bar{5}} = -G^{5\bar{4}} = G^{6\bar{7}} = -G^{7\bar{6}}
  = -i\frac{1}{2f_\pi^2}e^{-4\Im(\theta)/3} +
  i\frac{1}{2f_\pi^2}e^{2\Im(\theta)/3},\\
  &G^{8\bar{8}} =
  \frac{3e^{i2\theta/3 + i\bar{\theta}/3}}{f_\pi^2(e^{i\theta} + 2e^{i\bar{\theta}})},
\end{align}
and the generators are $T_A = \frac{1}{\sqrt{2}}\lambda_A$, where
$\lambda_A$ are the Gell-Mann matrices. 

Let us decompose Eq.~\eqref{eq:BPS_SL3} into real and imaginary parts
\begin{align}
  \del_1a =
  -\frac{m e^{b/3}\left(\sin\frac{a}{3} + e^b
    \sin\frac{2a}{3}\right)}{f_\pi^2(1+2e^{2b})},\\
  \del_1b =
  -\frac{m e^{b/3}\left(\cos\frac{a}{3} - e^b\cos\frac{2a}{3}\right)}{f_\pi^2(1+2e^{2b})},
\end{align}
where we have defined the complex function $\theta = a+i b$, in terms
of two real-valued functions.
Notice that the only fixed points (vacua) of this system is $a=2\pi n$
and $b=0$ with $n\in\mathbb{Z}$.
If we consider the imaginary function, $b$, then around the vacuum
$a=2\pi$, the asymptotic behavior of $b$ when large and negative is
\beq
b \sim -3\log\frac{m x}{f_\pi^2} + {\rm const.},
\eeq
whereas if $b$ is large and positive, it goes as
\beq
b \sim \frac{3}{2}\log\frac{m x}{f_\pi^2} + {\rm const.}.
\eeq
Neither of these behaviors allow for $b$ to return to the vacuum
$b=0$.
This means that the system exhibits an instability such that when
$|b|$ is larger than some critical value, it cannot return to the
vacuum even if $a\simeq 2\pi$.
This, however, does not prove the absence 
of solutions to the equation \eqref{eq:BPS_SL3}.
We will leave this task to future studies.
We have nevertheless been seeking for numerical solutions without
finding any.

\subsubsection{$SL(N,\mathbb{C})$ case}

Here we generalize the previous section to $SL(N,\mathbb{C})$.
We first diagonalize an $SL(N,\mathbb{C})$ element $M$ as
\beq
&& M_0 = {\rm diag}\, \left(
\exp \left( i\frac{\theta}{N} \right),  \cdots ,
\exp \left( i\frac{\theta}{N} \right), 
\exp \left( -i\theta \frac{N-1}{N} \right) \right)
= \exp (i\theta T_0),    \label{eq:M0} \\
&& T_0 \equiv {\rm diag}\,
   \left(\frac{1}{N}, \cdots,  \frac{1}{N}, -\frac{N-1}{N}\right) .
\eeq
We consider the boundary conditions of $\theta$ 
for a domain wall:
$\theta=0$, ($M={\bf 1}_N$) at $x \to -\infty$ 
and $\theta=2\pi$, ($M=\omega {\bf 1}_N$) at $x \to +\infty$.

Substituting this form into 
the BPS equation (\ref{eq:BPS-M}),
we get
\beq
i\del_1\theta = \frac{m e^{\frac{i(N-1)\theta}{N}}
  \left(1 - e^{i\bar{\theta}}\right)}
{f_\pi^2\left(e^{i\theta} + (N-1)e^{i\bar{\theta}}\right)}.
\label{eq:BPS_SLN}
\eeq

Since, as we have seen in the previous section, it is difficult at
best to find solutions in the generic case where the quasi-NG bosons
are turned on, we will first attempt a simplification.
We want to take the large $N$ limit of the above equation.
Let us define
\beq
\tilde{m} \equiv \frac{m}{N f_\pi^2}.
\eeq
In the large $N$ limit, Eq.~\eqref{eq:BPS_SLN} reduces to
\beq
i\del_1\theta = \tilde{m} e^{i\theta} (e^{-i\bar{\theta}} - 1).
\eeq
The vacua are clearly $\theta=2\pi n$, with $n\in\mathbb{Z}$. 
A domain wall solution would thus go from $0$ to $2\pi$.
Let us again decompose the equation into real functions
\begin{align}
  \del_1a &= -
\tilde{m}
e^{-b}\sin a, \\
  \del_1b &= 
\tilde{m}
e^{-2b} (e^b\cos a - 1),
  \label{eq:largeNb}
\end{align}
where $\theta=a+i b$. 
Expanding Eq.~\eqref{eq:largeNb} in small $a$ yields
\beq
\del_1b &= 
\tilde{m}
e^{-2b}
\left[e^b\left(1-\frac{1}{2}a^2+\mathcal{O}(a^4)\right) - 1\right],
\eeq
which for $b=0$ and $a$ small but positive will drive $b$ negative.
It is easy to see from the right-hand side of \eqref{eq:largeNb} that
once $b$ is negative, it will always decrease and hence become more
and more negative.
Since all the vacua has $b=0$, no solution exists to this equation.

Since we have used a particular -- albeit well motivated -- Ansatz for
the domain wall field $M$ and we have taken the large $N$ limit, this
is not a general proof of non-existence.

Finally, let us consider the finite $N$ case.
Decomposing Eq.~\eqref{eq:BPS_SLN} into real functions, we get
\begin{align}
  \del_1 a &= - 
\frac{m}{f_{\pi}^2}
\frac{e^{\frac{b}{N}}\left(\sin\frac{a}{N} +
  e^b\sin\frac{(N-1)a}{N}\right)}{1 + (N-1)e^{2b}}, \\
  \del_1 b &= - 
\frac{m}{f_{\pi}^2}
\frac{e^{\frac{b}{N}}\left(\cos\frac{a}{N} -
  e^b\cos\frac{(N-1)a}{N}\right)}{1 + (N-1)e^{2b}}, \label{eq:Nb}
\end{align}
where $\theta=a+i b$.
Expanding Eq.~\eqref{eq:Nb} in small $a$ yields
\beq
  \del_1 b = - 
\frac{m}{f_{\pi}^2}
e^{\frac{b}{N}}\left(1 - \frac{a^2}{2N^2} -
  e^b\left(1 - \frac{(N-1)^2a^2}{2N^2} + \mathcal{O}(a^4)\right)
   + \mathcal{O}(a^4)\right).
\eeq
Since the $SU(2)$ case is already solved, we will consider only $N>2$,
in which case the second cosine dominates and hence for $b=0$ and
small $a$ again drives $b$ negative.
If $b$ attains a negative value and it has to return to zero for when
$a$ goes to $2\pi$, then a positive value of the right-hand side of
Eq.~\eqref{eq:Nb} is a necessity.
The larger negative values $b$ takes on, the harder it is for the
function to be positive; therefore we will consider $b=0$ as the most
conservative choice for a negative value of $b$. If the function
cannot attain positive values for $b=0$, then even less so for $b<0$.
It is thus enough to realize that
\beq
-\cos\frac{a}{N} + \cos\frac{(N-1)a}{N} \leq 0,  \qquad {\rm for} \ \
N\geq 4.
\eeq
Hence, no solution exists for $N\geq 4$.
This is of course consistent with the large $N$ limit considered
above.
The only possibility is $N=3$ for which we do not have a proof at
present.
Numerically, however, we have not been able to find a solution to the
BPS equation. 

Of course the proof of non-existence is limited to the use of our
Ansatz. We leave a general proof for future work.

\subsection{Quadratic potential}

In this section we turn to the case of the quadratic potential
\eqref{eq:quadpot}, hence the BPS equation reads
\beq
 i M^{-1}\del_1 M = \frac{i m}{N^2} \tr(M^\dag) \tr(M^\dag T_{\bar{B}})
 T_A G^{A\bar{B}}.
\eeq
Restricting again to the NG-boson subspace, $M=M|_{\sigma^i=0}=U$, we
have
\beq
 i U^\dag\del_1 U = \frac{i m}{N^2 f_\pi^2} \tr(U^\dag) \tr(U^\dag T_A)
 T_A.
\eeq

With this superpotential, we can also calculate the domain wall
tension using Eq.~\eqref{eq:DWtension}, which for a domain wall
between $M=\mathbf{1}_N$ and the new vacuum yields
\beq
T = \frac{m}{N^2}\left|\Re\left((\tr\mathbf{1}_N)^2\right)\right| = m,
\label{eq:DWT_oldnew}
\eeq
while for a domain wall between the vacuum $M=\omega^k\mathbf{1}_N$
and the new vacuum, we have
\beq
T =
\frac{m}{N^2}\left|\Re\left((\omega^k\tr\mathbf{1}_N)^2\right)\right|
= m \cos\frac{4\pi k}{N}.
\eeq
If we restrict to the NG subspace, $M=M|_{\sigma^i=0}=U$, only real
vacua exist and so the tension is always given by
Eq.~\eqref{eq:DWT_oldnew}.

\subsubsection{$SU(2)$ solution}

Let us consider $N=2$ as a warm up.
The old vacua have $\omega = e^{i\pi}$ and so are given by
\beq
M_1 = \mathbf{1}_2,\qquad
M_2 = -\mathbf{1}_2,
\eeq
whereas the new vacuum is given by
\beq
\tr M = 0,
\eeq
which we can flesh out as
\beq
M_3 = 
\begin{pmatrix}
  a & b\\
  c & -a
\end{pmatrix},
\eeq
whose determinant is
\beq
-a^2 - b c = 1,
\eeq
yielding
\beq
M_3 = 
\begin{pmatrix}
  a & -\frac{1+a^2}{c}\\
  c & -a
\end{pmatrix}.
\eeq
The simplest possibility is $a=0$ and $c=1$, i.e.,
\beq
M_3 = -i\tau^2.
\eeq
The complication of the $N=2$ case is that there is no new diagonal
vacuum.
We will now consider an Ansatz that will interpolate between one of
the old vacua and the new vacuum, namely from $M_1$ to $M_3$:
\beq
U = \mathbf{1}_2 \cos\theta - i\tau^2 \sin\theta.
\label{eq:U_SU2_quadpot}
\eeq
Since both vacua are in the subspace spanned by the NG bosons, it is
consistent to restrict to the NG submanifold, if a solution exists.
The boundary conditions are $\theta=0$ ($U=\mathbf{1}_2$) at
$x\to +\infty$ and $\theta=\pi/2$ ($U=-i\tau^2$) at $x\to -\infty$.
Notice that due to the two vacua not being proportional to the
identity matrix ($\mathbf{1}_2$), the global $SU(2)$ symmetry is
broken to $U(1)$ by the vacuum.
Plugging the above Ansatz into Eq.~\eqref{eq:BPS-U} we get
\beq
\tau^2 \del_1\theta = -\tau^2 \frac{m}{4f_\pi^2} \sin 2\theta,
\label{eq:BPS_SU2_quadpot}
\eeq
which has the solution
\beq
\theta(x^1) = \arctan\exp\left[-\frac{m}{2f_\pi^2}(x^1 - X)\right],
\label{eq:BPS_SU2_quadpot_sol}
\eeq
where $X$ is again a position modulus. 

The $U(1)$ symmetry possessed by the vacuum is unbroken by the domain
wall solution. 
Consequently, the domain wall has no orientational moduli.

\subsubsection{$SU(2K)$ solutions}

We can straightforwardly extend the $SU(2)$ solution to $SU(2K)$, by
embedding $K$ blocks of the $SU(2)$ Ansatz \eqref{eq:U_SU2_quadpot} in
$U_0$ as
\beq
U_0 =
\begin{pmatrix}
  \mathbf{1}_2\cos\theta_1 - i\tau^2\sin\theta_1 \\
  &\ddots\\
  && \mathbf{1}_2\cos\theta_K - i\tau^2\sin\theta_K
\end{pmatrix},
\eeq
which interpolates between the vacuum $U=\mathbf{1}_{2K}$ and
\beq
U =
\begin{pmatrix}
  0 & -1\\
  1 & 0\\
  &&\ddots\\
  &&& 0 & -1\\
  &&& 1 & 0
\end{pmatrix}.
\eeq
It is straightforward to show that the BPS equation is exactly the
same as \eqref{eq:BPS_SU2_quadpot} in each $K$ block along the
diagonal.
The solution is therefore Eq.~\eqref{eq:BPS_SU2_quadpot_sol},
\beq
\theta_i(x^1) = \arctan\exp\left[-\frac{m}{2f_\pi^2}(x^1 - X_i)\right],
\eeq
with $i=1,\ldots,K$ and the moduli space is now given by
\beq
\frac{SU(K)}{U(1)^{K-1}},
\eeq
for generic position moduli $X_1\neq X_2\neq\cdots\neq X_K$.
If however $X_1=X_2=\cdots=X_K$ then no orientational (NG) moduli
exist for this solution.

\section{Low-energy effective theory on the domain
  wall}\label{sec:eff_th}

In this section, we construct the low-energy effective theory on the 
$SU(2)$ domain wall for the linear superpotential \eqref{eq:linearpot}
by using the moduli (or Manton's) approximation \cite{Manton:1981mp}.
The most general solution is obtained by performing the
$SU(2)_{\text{L} + \text{R}}$ transformation in
Eq.~\eqref{eq:SU2_solution} and is given by  
\begin{align}
U = V U_0 V^{\dagger} = \exp (i \theta V T_0 V^{\dagger}), \quad V \in SU(2).
\end{align}
Now we define the complex $2$-vector $\phi$ by the following relation
\begin{align}
V T_0 V^{\dagger} = \phi \phi^{\dagger} - \frac{1}{2} \mathbf{1}_2.
\end{align}
The vector $\phi$ satisfies the constraint $\phi^{\dagger} \phi = 1$.
Using this vector, the general solution is rewritten as
\begin{align}
U = \exp\left[
i \theta (\phi \phi^{\dagger} - \frac{1}{2} \mathbf{1}_2)
\right]
= 
\left[
\mathbf{1}_2 + (e^{i\theta} - 1) \phi \phi^{\dagger}
\right]
\exp(-i \theta/2).
\end{align}
The vector $\phi$ parametrizes $\mathbb{C}P^1$ and the moduli of
the solution are given by $X$ and $\phi$.

We first promote the moduli $X$ and $\phi$
in the solution to 
fields $X(x^\alpha)$ and $\phi(x^\alpha)$ 
depending on the coordinates $x^\alpha$ of the 
domain wall world-volume, 
substitute it into the original Lagrangian,
and then perform an integration over the codimension.

The differentiation of $U$ with respect to the world-volume 
coordinates $x^\alpha$ can be calculated as 
\begin{align}
\del_\alpha U &= 
\bigg[
-
\frac{i}{2}
\left({\bf 1}_2 + (e^{i\theta} -1) \phi \phi^\dagger\right)
  \del_\alpha \theta 
+  i  \del_\alpha \theta e^{i\theta} \phi \phi^\dagger \non
&\phantom{=\bigg[\ }
+ (e^{i\theta} -1) 
 (\del_\alpha \phi \phi^\dagger
 + \phi  \del_\alpha \phi^\dagger)
  \bigg] \exp (
-i\theta/2
) .  
\end{align}
By using the relations
\begin{align}
\del_\alpha e^{i\theta(x^1;X(x^\alpha))}
&= i \del_\alpha X \frac{\del \theta}{\del X} e^{i\theta} = - i \del_\alpha X 
 \del_{1} \theta e^{i\theta},\\
2|1-e^{i\theta}|^2 &= 2(2 - e^{i\theta} - e^{-i\theta}) = 8\sin^2 \frac{\theta}{2},
\end{align}
we obtain
\beq
\tr\left(\del_\alpha U \del^\alpha U^\dagger\right) 
= 
 \frac{1}{2} 
(\del_1 \theta)^2 (\del_\alpha X)^2 +8 \sin^2\frac{\theta}{2}  
 \left[\del^\alpha \phi^\dagger \del_\alpha \phi 
       + (\phi^\dagger \del_\alpha \phi)^2 \right]. \label{eq:eff_th}
\eeq
By noting the formulas
\beq
 \int dx^1   \sin^2\frac{\theta}{2}  = \frac{2}{m}, 
\qquad
 \int dx^1 \;(\del_1\theta)^2 
 = \int dx^1 \left(\frac{m}{f_\pi^2} \sin\frac{\theta}{2}\right)^2
 = \frac{2m}{f_\pi^4},
\eeq
where we have used the BPS equation (\ref{eq:SG})
in the second relation, 
the integration of Eq.~(\ref{eq:eff_th}) over 
the codimensional coordinate $x$ 
yields the final form of the effective Lagrangian on the wall:
\beq
{\cal L}_{\rm eff} 
= 
- \frac{m}{f_\pi^2}
  \del_\alpha X \del^\alpha X 
-
\frac{16 f_\pi^2}{m}
\left[
\del_\alpha \phi^\dagger \del^\alpha \phi + (\phi^\dagger\del_\alpha \phi)(\phi^\dagger\del^\alpha \phi)
\right]
-\frac{m}{f_\pi^2},
\eeq
where the last term is the tension of the domain wall. 
The first term describes the translational zero modes while the second term  
 represents the orientational zero modes,
 which is described by the $\mathbb{C}P^1$ model.

\section{Summary and discussion} \label{sec:conc}

We have studied the BPS domain walls in the $\mathcal{N}=1$
supersymmetric chiral Lagrangian with $SU(N)_{\rm L+R}$ invariant pion
mass terms.
The bosonic components of the model consist of both NG and quasi-NG
bosons.  
We have constructed exact solutions of BPS pion domain walls
for the case of a linear and a quadratic superpotential.
In all cases we have considered the simplest K\"ahler potential; the
difference with the most general K\"ahler potential of the chiral
invariant amounts simply to a change in the K\"ahler modulus (pion
decay constant).
All the domain wall solutions are topologically stable.
We have, however, found not all vacua are connected by domain walls of
BPS type.
In particular, we have only been able to find BPS domain wall
solutions connecting pion vacua, i.e.~vacua with no imaginary part. 
These domain wall solutions, in turn, are described only by the
NG-boson subspace and not by the quasi-NG bosons, which are left
turned off in the solutions.
For a well-motivated Ansatz, we have found the complex BPS equation
not restricted to the NG submanifold.
We have, however, proved that this BPS equation has no solutions for
$N\geq 4$. The $N=2$ case has only real vacua and the analytic domain
wall solution is simply the sine-Gordon solution.
We have not been able to find analytical or numerical solutions to the
BPS equation for $N=3$, although we do not at present have a proof of
non-existence.
The understanding of the absence of domain walls between all the vacua
with an imaginary part still needs some progress. This may in turn
teach us about the dynamics of the quasi-NG bosons in nonperturbative
solutions. We leave this interesting open issue for future studies. 

The BPS bound gives a tension which is the absolute value of the
real part of the difference between the superpotential evaluated at
two given vacua.
The fact that the tension is the real part of this difference, means
that if the vacua are purely imaginary (for instance $M=i$ and
$M=-i$), then the BPS bound gives a vanishing tension. Since,
physically, no domain wall can interpolate two such vacua with
vanishing tension, they are necessarily not saturating the bound and
thus are non-BPS.
Whether non-BPS solutions exist or not is beyond the scope of this
paper, although it is an interesting problem which we leave for future
work.

Non-Abelian vortices in $U(2)$ gauge theories 
also carry ${\mathbb C}P^{1}$ moduli 
\cite{Hanany:2003hp} and the $U(N)$ gauge group was
generalized to an arbitrary gauge group \cite{Eto:2008yi} 
such as $SO(N)$ and $USp(2N)$ \cite{Eto:2009bg}.
Our model itself could straightforwardly be extended to a chiral
Lagrangian of an arbitrary group $G$, but 
one nontrivial question is which coset space $G/H$ 
is realized on the domain wall. 
We have already observed a more complicated structure of the
domain walls in our model than simply the $\mathbb{C}P^{N-1}$ model;
further cosets appear already in the $SU(N)$ case.

Our model should admit a domain wall junction 
as a 1/4 BPS state \cite{Gibbons:1999np}.
In particular, 
the simplest superpotential with $N$ vacua 
is expected to admit a ${\mathbb Z}_N$ symmetric 
domain wall junction 
as in Ref.~\cite{Oda:1999az}.
However, since we already have observed that not all the vacua are
connected in our model, the domain wall junctions may be either absent
or modified compared to the usual case.

The effective theory
of a non-Abelian vortex in $U(N)$ gauge theory 
is the ${\mathbb C}P^{N-1}$ model,
and lump solutions on it correspond to
Yang-Mills instantons in the bulk \cite{Hanany:2004ea}.
The total configuration of lumps inside a vortex 
is a 1/4 BPS state. 
In the same way, lump solutions in 
our domain wall, 
which will correspond to Skyrmions in the bulk as the case 
of a non-Abelian sine-Gordon soliton \cite{Eto:2015uqa}, 
may be 1/4 BPS states.

Recently, a supersymmetric Skyrme term
has been constructed in Ref.~\cite{Gudnason:2015ryh}
in which it has been found that the 
usual kinetic term cancels out. 
In this case, the introduction of a superpotential 
can be done only perturbatively \cite{Adam:2013awa}.
Construction of such a model and 
its BPS domain wall solutions -- that may be of compacton type --
remain as a future problem.

The chiral Lagrangian can be realized 
on a non-Abelian domain wall 
in ${\cal N}=2$ supersymmetric $U(N)$ gauge theory
with two $N\times N$ complex scalar fields 
(hypermultiplets)  
\cite{Shifman:2003uh,Eto:2005cc}. 
If we find a suitable mass deformation preserving 
(part of) the supersymmetry 
in the original bulk action that induces 
the superpotential $W=\frac{m}{N}\tr (M)$ 
on the wall, then 
our solution may describe 
a wall inside a wall as a 1/4 BPS state.
The non-Abelian domain wall in 
Refs.~\cite{Shifman:2003uh,Eto:2005cc} 
describes a non-Abelian Josephson junction 
in the presence of 
a Josephson term in the bulk that breaks supersymmetry, 
and a sine-Gordon soliton on the wall that 
describes a non-Abelian vortex absorbed 
into the junction 
\cite{Nitta:2015mma}.
A supersymmetry-preserving mass deformation, 
if it exists, 
would describe a supersymmetric Josephson junction 
and would 
give a BPS non-Abelian vortex 
absorbed into the junction, 
that is, a BPS non-Abelian Josephson vortex.

Finally,
in (non-supersymmetric) QCD, 
topological solitons in chiral symmetry breaking 
were studied, see, e.~g.~Refs.~\cite{Eto:2013hoa,Eto:2013bxa}.
Our BPS configurations may have implications 
for these more realistic cases as well.

\subsection*{Acknowledgments}

S.~B.~G.~thanks the Recruitment Program of High-end Foreign
Experts for support.
The work of M.~N.~is supported in part by a Grant-in-Aid for
Scientific Research on Innovative Areas ``Topological Materials
Science'' (KAKENHI Grant No.~15H05855) and ``Nuclear Matter in Neutron
Stars Investigated by Experiments and Astronomical Observations''
(KAKENHI Grant No.~15H00841) from the the Ministry of Education,
Culture, Sports, Science (MEXT) of Japan. The work of M.~N.~is also
supported in part by the Japan Society for the Promotion of Science
(JSPS) Grant-in-Aid for Scientific Research (KAKENHI Grant
No.~25400268) and by the MEXT-Supported Program for the Strategic
Research Foundation at Private Universities ``Topological Science''
(Grant No.~S1511006).
The work of S. S. is supported in part by Kitasato University
Research Grant for Young Researchers.


\end{document}